\documentclass{article}

    \usepackage{graphicx} 
    \usepackage{adjustbox} 
    \usepackage{color} 
    \usepackage{enumerate} 
    \usepackage{geometry} 
    \usepackage{amsmath} 
    \usepackage{amssymb} 
    \usepackage[mathletters]{ucs} 
    \usepackage[utf8x]{inputenc} 
    \usepackage{fancyvrb} 
    \usepackage{grffile} 
    \usepackage{hyperref}
    \usepackage{authblk}
    \usepackage{longtable} 
    \usepackage{booktabs}  
    
\usepackage{mathtools}
\usepackage{soul}

\usepackage[format=plain, labelfont={it}, textfont=it]{caption}

    \author[1]{Pierre Cohort\thanks{pcohort@zeliade.com}}
    \author[1]{Jacopo Corbetta\thanks{jacorbetta@zeliade.com}}
    \author[1]{Ismail Laachir\thanks{ilaachir@zeliade.com, Corresponding author}}
      
    \affil[1]{Zeliade Systems, 56 rue Jean-Jacques Rousseau, Paris, France}

    \definecolor{orange}{cmyk}{0,0.4,0.8,0.2}
    \definecolor{darkorange}{rgb}{.71,0.21,0.01}
    \definecolor{darkgreen}{rgb}{.12,.54,.11}
    \definecolor{myteal}{rgb}{.26, .44, .56}
    \definecolor{gray}{gray}{0.45}
    \definecolor{lightgray}{gray}{.95}
    \definecolor{mediumgray}{gray}{.8}
    \definecolor{inputbackground}{rgb}{.95, .95, .85}
    \definecolor{outputbackground}{rgb}{.95, .95, .95}
    \definecolor{traceback}{rgb}{1, .95, .95}
    \definecolor{red}{rgb}{.6,0,0}
    \definecolor{green}{rgb}{0,.65,0}
    \definecolor{brown}{rgb}{0.6,0.6,0}
    \definecolor{blue}{rgb}{0,.145,.698}
    \definecolor{purple}{rgb}{.698,.145,.698}
    \definecolor{cyan}{rgb}{0,.698,.698}
    \definecolor{lightgray}{gray}{0.5}
    
    \definecolor{darkgray}{gray}{0.25}
    \definecolor{lightred}{rgb}{1.0,0.39,0.28}
    \definecolor{lightgreen}{rgb}{0.48,0.99,0.0}
    \definecolor{lightblue}{rgb}{0.53,0.81,0.92}
    \definecolor{lightpurple}{rgb}{0.87,0.63,0.87}
    \definecolor{lightcyan}{rgb}{0.5,1.0,0.83}
    
    \DefineVerbatimEnvironment{Highlighting}{Verbatim}{commandchars=\\\{\}}

    \title{Analytical scores for stress scenarios}

\makeatletter
\def\PY@reset{\let\PY@it=\relax \let\PY@bf=\relax%
    \let\PY@ul=\relax \let\PY@tc=\relax%
    \let\PY@bc=\relax \let\PY@ff=\relax}
\def\PY@tok#1{\csname PY@tok@#1\endcsname}
\def\PY@toks#1+{\ifx\relax#1\empty\else%
    \PY@tok{#1}\expandafter\PY@toks\fi}
\def\PY@do#1{\PY@bc{\PY@tc{\PY@ul{%
    \PY@it{\PY@bf{\PY@ff{#1}}}}}}}
\def\PY#1#2{\PY@reset\PY@toks#1+\relax+\PY@do{#2}}

\expandafter\def\csname PY@tok@gd\endcsname{\def\PY@tc##1{\textcolor[rgb]{0.63,0.00,0.00}{##1}}}
\expandafter\def\csname PY@tok@gu\endcsname{\let\PY@bf=\textbf\def\PY@tc##1{\textcolor[rgb]{0.50,0.00,0.50}{##1}}}
\expandafter\def\csname PY@tok@gt\endcsname{\def\PY@tc##1{\textcolor[rgb]{0.00,0.27,0.87}{##1}}}
\expandafter\def\csname PY@tok@gs\endcsname{\let\PY@bf=\textbf}
\expandafter\def\csname PY@tok@gr\endcsname{\def\PY@tc##1{\textcolor[rgb]{1.00,0.00,0.00}{##1}}}
\expandafter\def\csname PY@tok@cm\endcsname{\let\PY@it=\textit\def\PY@tc##1{\textcolor[rgb]{0.25,0.50,0.50}{##1}}}
\expandafter\def\csname PY@tok@vg\endcsname{\def\PY@tc##1{\textcolor[rgb]{0.10,0.09,0.49}{##1}}}
\expandafter\def\csname PY@tok@m\endcsname{\def\PY@tc##1{\textcolor[rgb]{0.40,0.40,0.40}{##1}}}
\expandafter\def\csname PY@tok@mh\endcsname{\def\PY@tc##1{\textcolor[rgb]{0.40,0.40,0.40}{##1}}}
\expandafter\def\csname PY@tok@go\endcsname{\def\PY@tc##1{\textcolor[rgb]{0.53,0.53,0.53}{##1}}}
\expandafter\def\csname PY@tok@ge\endcsname{\let\PY@it=\textit}
\expandafter\def\csname PY@tok@vc\endcsname{\def\PY@tc##1{\textcolor[rgb]{0.10,0.09,0.49}{##1}}}
\expandafter\def\csname PY@tok@il\endcsname{\def\PY@tc##1{\textcolor[rgb]{0.40,0.40,0.40}{##1}}}
\expandafter\def\csname PY@tok@cs\endcsname{\let\PY@it=\textit\def\PY@tc##1{\textcolor[rgb]{0.25,0.50,0.50}{##1}}}
\expandafter\def\csname PY@tok@cp\endcsname{\def\PY@tc##1{\textcolor[rgb]{0.74,0.48,0.00}{##1}}}
\expandafter\def\csname PY@tok@gi\endcsname{\def\PY@tc##1{\textcolor[rgb]{0.00,0.63,0.00}{##1}}}
\expandafter\def\csname PY@tok@gh\endcsname{\let\PY@bf=\textbf\def\PY@tc##1{\textcolor[rgb]{0.00,0.00,0.50}{##1}}}
\expandafter\def\csname PY@tok@ni\endcsname{\let\PY@bf=\textbf\def\PY@tc##1{\textcolor[rgb]{0.60,0.60,0.60}{##1}}}
\expandafter\def\csname PY@tok@nl\endcsname{\def\PY@tc##1{\textcolor[rgb]{0.63,0.63,0.00}{##1}}}
\expandafter\def\csname PY@tok@nn\endcsname{\let\PY@bf=\textbf\def\PY@tc##1{\textcolor[rgb]{0.00,0.00,1.00}{##1}}}
\expandafter\def\csname PY@tok@no\endcsname{\def\PY@tc##1{\textcolor[rgb]{0.53,0.00,0.00}{##1}}}
\expandafter\def\csname PY@tok@na\endcsname{\def\PY@tc##1{\textcolor[rgb]{0.49,0.56,0.16}{##1}}}
\expandafter\def\csname PY@tok@nb\endcsname{\def\PY@tc##1{\textcolor[rgb]{0.00,0.50,0.00}{##1}}}
\expandafter\def\csname PY@tok@nc\endcsname{\let\PY@bf=\textbf\def\PY@tc##1{\textcolor[rgb]{0.00,0.00,1.00}{##1}}}
\expandafter\def\csname PY@tok@nd\endcsname{\def\PY@tc##1{\textcolor[rgb]{0.67,0.13,1.00}{##1}}}
\expandafter\def\csname PY@tok@ne\endcsname{\let\PY@bf=\textbf\def\PY@tc##1{\textcolor[rgb]{0.82,0.25,0.23}{##1}}}
\expandafter\def\csname PY@tok@nf\endcsname{\def\PY@tc##1{\textcolor[rgb]{0.00,0.00,1.00}{##1}}}
\expandafter\def\csname PY@tok@si\endcsname{\let\PY@bf=\textbf\def\PY@tc##1{\textcolor[rgb]{0.73,0.40,0.53}{##1}}}
\expandafter\def\csname PY@tok@s2\endcsname{\def\PY@tc##1{\textcolor[rgb]{0.73,0.13,0.13}{##1}}}
\expandafter\def\csname PY@tok@vi\endcsname{\def\PY@tc##1{\textcolor[rgb]{0.10,0.09,0.49}{##1}}}
\expandafter\def\csname PY@tok@nt\endcsname{\let\PY@bf=\textbf\def\PY@tc##1{\textcolor[rgb]{0.00,0.50,0.00}{##1}}}
\expandafter\def\csname PY@tok@nv\endcsname{\def\PY@tc##1{\textcolor[rgb]{0.10,0.09,0.49}{##1}}}
\expandafter\def\csname PY@tok@s1\endcsname{\def\PY@tc##1{\textcolor[rgb]{0.73,0.13,0.13}{##1}}}
\expandafter\def\csname PY@tok@kd\endcsname{\let\PY@bf=\textbf\def\PY@tc##1{\textcolor[rgb]{0.00,0.50,0.00}{##1}}}
\expandafter\def\csname PY@tok@sh\endcsname{\def\PY@tc##1{\textcolor[rgb]{0.73,0.13,0.13}{##1}}}
\expandafter\def\csname PY@tok@sc\endcsname{\def\PY@tc##1{\textcolor[rgb]{0.73,0.13,0.13}{##1}}}
\expandafter\def\csname PY@tok@sx\endcsname{\def\PY@tc##1{\textcolor[rgb]{0.00,0.50,0.00}{##1}}}
\expandafter\def\csname PY@tok@bp\endcsname{\def\PY@tc##1{\textcolor[rgb]{0.00,0.50,0.00}{##1}}}
\expandafter\def\csname PY@tok@c1\endcsname{\let\PY@it=\textit\def\PY@tc##1{\textcolor[rgb]{0.25,0.50,0.50}{##1}}}
\expandafter\def\csname PY@tok@kc\endcsname{\let\PY@bf=\textbf\def\PY@tc##1{\textcolor[rgb]{0.00,0.50,0.00}{##1}}}
\expandafter\def\csname PY@tok@c\endcsname{\let\PY@it=\textit\def\PY@tc##1{\textcolor[rgb]{0.25,0.50,0.50}{##1}}}
\expandafter\def\csname PY@tok@mf\endcsname{\def\PY@tc##1{\textcolor[rgb]{0.40,0.40,0.40}{##1}}}
\expandafter\def\csname PY@tok@err\endcsname{\def\PY@bc##1{\setlength{\fboxsep}{0pt}\fcolorbox[rgb]{1.00,0.00,0.00}{1,1,1}{\strut ##1}}}
\expandafter\def\csname PY@tok@mb\endcsname{\def\PY@tc##1{\textcolor[rgb]{0.40,0.40,0.40}{##1}}}
\expandafter\def\csname PY@tok@ss\endcsname{\def\PY@tc##1{\textcolor[rgb]{0.10,0.09,0.49}{##1}}}
\expandafter\def\csname PY@tok@sr\endcsname{\def\PY@tc##1{\textcolor[rgb]{0.73,0.40,0.53}{##1}}}
\expandafter\def\csname PY@tok@mo\endcsname{\def\PY@tc##1{\textcolor[rgb]{0.40,0.40,0.40}{##1}}}
\expandafter\def\csname PY@tok@kn\endcsname{\let\PY@bf=\textbf\def\PY@tc##1{\textcolor[rgb]{0.00,0.50,0.00}{##1}}}
\expandafter\def\csname PY@tok@mi\endcsname{\def\PY@tc##1{\textcolor[rgb]{0.40,0.40,0.40}{##1}}}
\expandafter\def\csname PY@tok@gp\endcsname{\let\PY@bf=\textbf\def\PY@tc##1{\textcolor[rgb]{0.00,0.00,0.50}{##1}}}
\expandafter\def\csname PY@tok@o\endcsname{\def\PY@tc##1{\textcolor[rgb]{0.40,0.40,0.40}{##1}}}
\expandafter\def\csname PY@tok@kr\endcsname{\let\PY@bf=\textbf\def\PY@tc##1{\textcolor[rgb]{0.00,0.50,0.00}{##1}}}
\expandafter\def\csname PY@tok@s\endcsname{\def\PY@tc##1{\textcolor[rgb]{0.73,0.13,0.13}{##1}}}
\expandafter\def\csname PY@tok@kp\endcsname{\def\PY@tc##1{\textcolor[rgb]{0.00,0.50,0.00}{##1}}}
\expandafter\def\csname PY@tok@w\endcsname{\def\PY@tc##1{\textcolor[rgb]{0.73,0.73,0.73}{##1}}}
\expandafter\def\csname PY@tok@kt\endcsname{\def\PY@tc##1{\textcolor[rgb]{0.69,0.00,0.25}{##1}}}
\expandafter\def\csname PY@tok@ow\endcsname{\let\PY@bf=\textbf\def\PY@tc##1{\textcolor[rgb]{0.67,0.13,1.00}{##1}}}
\expandafter\def\csname PY@tok@sb\endcsname{\def\PY@tc##1{\textcolor[rgb]{0.73,0.13,0.13}{##1}}}
\expandafter\def\csname PY@tok@k\endcsname{\let\PY@bf=\textbf\def\PY@tc##1{\textcolor[rgb]{0.00,0.50,0.00}{##1}}}
\expandafter\def\csname PY@tok@se\endcsname{\let\PY@bf=\textbf\def\PY@tc##1{\textcolor[rgb]{0.73,0.40,0.13}{##1}}}
\expandafter\def\csname PY@tok@sd\endcsname{\let\PY@it=\textit\def\PY@tc##1{\textcolor[rgb]{0.73,0.13,0.13}{##1}}}

\makeatother

    \definecolor{incolor}{rgb}{0.0, 0.0, 0.5}
    \definecolor{outcolor}{rgb}{0.545, 0.0, 0.0}

    \hypersetup{
      breaklinks=true,  
      colorlinks=true,
      urlcolor=blue,
      linkcolor=darkorange,
      citecolor=darkgreen,
      }
    
\begin{document}

     \hypersetup{ breaklinks=true,  
      colorlinks=true,
      urlcolor=blue,
      linkcolor=darkorange,
      citecolor=darkgreen,      
      }

\begin{figure}
\hfill\includegraphics[height=0.1\textwidth]{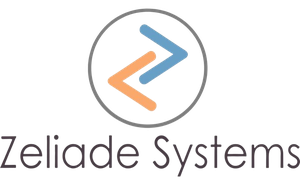}
\end{figure}
\maketitle

\begin{abstract}
	In this work, inspired by the Archer-Mouy-Selmi approach
(\cite{MouyArcherSelmi}), we present two methodologies for scoring the
stress test scenarios used by CCPs for sizing their Default Funds. These
methodologies can be used by risk managers to compare different sets of
scenarios and could be particularly useful when evaluating the relevance
of adding new scenarios to a pre-existing set.
\end{abstract}

\section{Comparison of sets of Hypothetical stress
scenarios}\label{comparison-of-sets-of-hypothetical-stress-scenarios}

	After the financial crisis of 2008, the topic of stress testing got more
and more attention from the financial infrastructures environment,
specifically from Central Clearing Counterparties (CCPs). Both the
PFMI-IOSCO \cite{PFMI_IOSCO} and EMIR regulations \cite{EMIR},
\cite{RTS_EMIR} require CCPs to specify \emph{extreme but plausible
scenarios} for the sizing of their default funds. Moreover the EMIR
regulation requires a review of the same stress scenarios (at least
annually according to Article 31 of EMIR RTS \cite{RTS_EMIR} and
according to the key explanations of Principles 4, 5, and 7 of PFMI
\cite{PFMI_IOSCO}). As such a risk manager may face the conundrum of
assessing the benefits of adding or changing sets of scenarios. In this
work we provide two methodologies giving a quantitative assessment of
the advantages or disadvantages of the new set of scenarios.

	\subsection{Plausible Hypothetical
Scenarios}\label{plausible-hypothetical-scenarios}

	A hypothetical scenario is, as the name suggests, a stylized scenario
designed in order to capture a tail risk. Hypothetical stress scenarios
can be designed starting from risk manager views, as for example a
parallel shift of all yield curves for fixed income products, or using
as a base a scenario obtained via quantitative methods, coming from the
fit of a distribution, or a PCA.

	By construction, the hypothetical scenarios do not refer directly to the
history. So, it is a good indicator of the scenarios quality to estimate
their plausibility. Moreover, the size of the hypothetical scenarios are
generally calibrated separately for each risk factor, so that the
plausibility of the joint scenarios does not have the same confidence
level as the one dimentional quantile level. The plausibility can be
estimated, for instance, by evaluating the log-likelihood of the
hypothetical scenarios, using a joint-distribution calibrated on the
risk factors. A joint Student distribution can be calibrated on the
historical data of the risk factors, then used to estimate the
likelihood of the scenario.

The problem of the design of extreme but plausible scenarios has been
tackled in the literature, for example by Thomas Breuer and co-authors
in \cite{breuer2001plausible}. They look for the scenario that gives the
worst loss, under a plausibility constraint (likelihood less than a
cap). While this approach is valid for portfolio management, it is not
well suited for the Stress Testing context of CCPs. In their second
paper \cite{breuer2008overcoming}, the authors consider a dual problem
whose solution does not depend on additional dimensionality of the
problem and which closely resembles the problems faced by a risk manager
at a CCP. This approach is made even more explicit by Q.Archer, P.Mouy
and M.Selmi (LCH), who proposed (cf. \cite{MouyArcherSelmi}) a framework
for the design of extreme but plausible scenarios. Their methodology
considers a linear portfolio $P$, so that the P\&Ls from risk factor
returns $s$ is simply $P^t s$, and assumes a calibrated historical
distribution of the risk factors, with density $f$.

	The approach can be formalized by the following maximum likelihood
problem:

\begin{equation}\label{Pb2}
 \max_{P^t s \leq q} f_{\theta}(\text{s})
\end{equation}

where

\begin{itemize}
\itemsep1pt\parskip0pt\parsep0pt
\item
  $s$ is the vector of the risk factors returns.
\item
  $P$ is the portfolio positions.
\item
  $f_{\theta}$ is the density function of the joint-distribution of all
  the risk factor returns $s$.
\item
  $q$ is a cap constraint on the portfolio loss.
\end{itemize}

	For instance a heavy-tail distribution, like a T- Student law can be
calibrated on the risk factors returns. $q$ can be chosen as the
$\alpha$-quantile of the distribution of the portfolio loss $P^t s$. For
example, if $s$ is a Gaussian vector, then $P^t s$ is Gaussian so that
$q$ can be chosen as a (one-dimensional) Gaussian quantile.

This methodology has the advantage of yielding extreme scenarios, with a
loss cap, so that we are ensured to have meaningful scenarios. Moreover,
if the distribution is chosen to be a standard elliptic (Student or
Gaussian, with correlation $\Sigma$, null average and marginal standard
deviation $1$), then the problem \eqref{Pb2} admits a simple closed
formula solution:

\begin{equation}\label{eq: sol_elliptic}
S^*(P) = q\frac{\Sigma P}{^t\! P\Sigma P}.
\end{equation}

	As pointed out in \cite{MouyArcherSelmi} and \cite{breuer2008overcoming}
another advantage of this methodology is the fact that, at least in an
elliptical setting, the solution of \eqref{Pb2} is not dependent on the
presence of additional risk factors not appearing in the portfolio,
while the primal problem considered in \cite{breuer2001plausible} gives
losses dependent on the introduction of risk factors unrelated to the
portfolio.

	\subsection{The score functions}\label{the-score-functions}

	We apply now the ideas from \cite{MouyArcherSelmi} to the comparison of
sets of stress scenarios.

We start by considering a single linear portfolio $P$ (such that the
P\&L associated to the risk factor return $s$ is $^t\!Ps$) and a set of
stress scenarios $\mathcal{S}:=\lbrace S_0,\ldots,S_n \rbrace$ for the
risk factor returns $s$.

We can thus calculate the stress loss associated to the portfolio $P$
as: \[
l(P) := \min_{0\le i\le n}{^t\!PS_i}\,,
\] we then select the scenario $\hat{S}(P)$ among $\mathcal{S}$ that
drive $l(P)$. If several drivers are found, we select the driver that
maximize the density function (another criterion could be applied, such
as minimizing the Mahalanobis distance
$^t\!\hat{S}(P)\Sigma^{-1}\hat{S}(P)$ among the candidate drivers).

	We then compute the scenario $S^*(P)$ solving

\begin{equation}\label{problem}
\sup_{S\,;\,{^t\!P}S\le{^t\!P}\hat{S}}f_{\theta}(S)\,,
\end{equation}

that is to say the most plausible scenario generating a loss equal (or
greater) to the worst loss obtained with the stress scenarios.
Equivalently, the two scenarios $\hat{S}(P)$ and $S^*(P)$ generate the
same loss $l(P)$ but $S^*(P)$ is the most plausible with respect to the
distribution assumption.

	We finally introduce the two score functions. The first score function
measures the quality of the ratio loss to plausibility of $\hat{S}(P)$,
and is given by

\begin{equation}\label{score}
\phi_{\mathcal{S}}(P):=\frac{f_{\theta}\left(\hat{S}(P)\right)}{f_{\theta}\left(S^*(P)\right)}\in \left]0,1\right]
\end{equation}

	The higher the score, the better it is, as a high $\phi$ indicates that
the stress scenarios contained in the set $\mathcal{S}$ are close, in a
plausibility sense, to the most likely scenario inducing the same level
of losses for the portfolio.

	The second score is a geometrical criterion, measuring to what extent a
driver is in the same direction as the optimal scenario.

\begin{equation}
\label{scorescal}
\psi_{\mathcal{S}}(P):=\frac{\langle \hat{S}(P), S^*(P)\rangle}{\left\|\hat{S}(P)\right\|\left\|S^*(P)\right\|}\in \left]-1,1\right]
\end{equation}

	Also in this case the higher the score, the better it is, as it
indicates that the ``risk direction'' of the portfolio is captured by
the stress scenario set $\mathcal{S}$.

	\subsection{Applying the scores to sets of
scenarios}\label{applying-the-scores-to-sets-of-scenarios}

	Suppose now that we have two sets of stress scenarios:
$\mathcal{S} = \{S_0, \cdots, S_n\},$and
$\mathcal{T} = \{T_0,\cdots, T_m\}$, possibly partially overlapping, and
we want to evaluate the advantages of one set with respect to the other.
The score functions we introduced in the previous section allows us to
do it in the following way:

\begin{itemize}
\itemsep1pt\parskip0pt\parsep0pt
\item
  Select a reference set of portfolios $P_0,\ldots,P_M$.
\item
  Calculate the values $\phi_{\mathcal{S}},\psi_{\mathcal{S}}(P_i)$'s
  and $\phi_{\mathcal{T}}, \psi_{\mathcal{T}}(P_i)$'s.
\item
  Define a final score from those values.
\end{itemize}

The choice of the final score depends on the risk manager view. In our
numerical result parts we propose two different approaches.

\begin{itemize}
\itemsep1pt\parskip0pt\parsep0pt
\item
  \textbf{Scenario approach}, which is particularly meaningful when the
  set $\mathcal{T}$ is a modification of the set $\mathcal{S}$. For each
  stress scenario we compute the average and standard deviation of the
  function $\psi$ and $\phi$ on the set of portfolios for which the
  scenario is the driver. This approach allows to have a view on which
  scenarios could be eventually modified, or even eliminated as being
  very far from optimality, either from a plausibility or geometrical
  points of view.
\item
  \textbf{Portfolio approach}. We compare the score functions on each
  portfolio. This approach allows to better understand for which
  portfolios the risk is not correctly sized, and it can be used for
  understanding which test portfolios are not sufficiently stressed by
  the current set of stress scenarios.
\end{itemize}

	We point out that the proposed scores should be used more as a non
rejection indicator and not as an acceptance one, similarly to the
Kupiec Test which says that an hypothesis can not be rejected, not that
it should be accepted.

	\subsection{Finding the optimal
scenario}\label{finding-the-optimal-scenario}

	For elliptical distributions, the solution of (\ref{problem}) can be
found exactly as described in \cite{MouyArcherSelmi}. However, for the
more generic meta-elliptical distributions (introduced in
\cite{fang2002meta}) this is no more the case. As these are the
distributions we will fit the risk factors returns on, we provide two
possible alternatives for finding the most plausible scenario at a given
loss.

We recall that a meta-elliptical distribution $f_{\theta}$ is a
multidimensional distribution with elliptical copula. The setting we
will consider consists of a T-Student copula with T-Student marginals,
and it is consequently characterized by a parameter $\theta^*$
containing:

\begin{itemize}
\itemsep1pt\parskip0pt\parsep0pt
\item
  the location vector $\mu$ and scale vector $\sigma$
\item
  the correlation matrix $\Sigma$
\item
  the vector of marginal degrees of freedom $\nu$
\item
  the degrees of freedom $\bar\nu$ of the copula(denoting also in the
  sequel the d-dimensional constant vector $(\bar\nu,\ldots,\bar\nu)$).
\end{itemize}

	\subsubsection{Approximate solution}\label{approximate-solution}

	The first method was proposed by Mouy et al. \cite{MouyArcherSelmi} and
it is based on approximating the meta-elliptical distribution by an
elliptical distribution, i.e.~using the same degree of freedom for the
copula and the marginals.

	We start by normalizing the distribution, via the linear transformation
$\tilde s:=(s-\mu)/\sigma$, and we get the equivalent problem

\[\sup_{\tilde s; {^t\!\tilde P}\tilde s\le \tilde q}\tilde f_{\theta^*}(\tilde s)\]

where

\begin{itemize}
\itemsep1pt\parskip0pt\parsep0pt
\item
  $\tilde f_{\theta^*}(\tilde s):=f_{\theta^*}(s)$
\item
  $\tilde P:=\sigma P$
\item
  $\tilde q:={^t\!P}(\hat{S}-\mu)$.
\end{itemize}

If the distribution $\tilde f_{\theta^*}$ was elliptical, the optimal
scenario for the problem above would be given by
\eqref{eq: sol_elliptic} \[
\tilde S^*(P) = \tilde q\frac{\Sigma \tilde P}{^t\!\tilde P\Sigma \tilde P}\,,
\] Transporting it back to the original problem, one has: \[
S^*(P):=\mu+\sigma \left(\tilde q\frac{\Sigma \tilde P}{^t\!\tilde P\Sigma \tilde P}\right)\,.
\]

	As stated above, the approximated solution is obtained by approximating
the meta-elliptical distribution with an elliptical distribution,
obtaining the sub-optimal scenario

\[
\bar S(P):=\mu+\sigma T_{\nu}^{-1}\circ T_{\bar\nu}\left(\tilde q\frac{\Sigma \tilde P}{^t\!\tilde P\Sigma \tilde P}\right)
\]

where $T_{\nu}(x)$ is the vector $(T_{\nu_i}(x_i))_{1\leq i\leq d}$,
$T_{\nu_i}$ being the CDF of a standard T-Student distribution with
$\nu_i$ degrees of freedom.

We point out that the approximation quality is strongly linked to the
``almost linearity'' of the function $T_{\nu}^{-1}\circ T_{\bar\nu}$
around $0$. In the case where $\bar\nu$ and $\nu$ are very different,
the approximation could be poor, with significant discrepancies both in
term of optimal density value and on loss constraint violation.

	\subsubsection{Exact numerical solution}\label{exact-numerical-solution}

	An exact solution can also be recovered numerically using classical
optimizers. In fact the target function is easy and fast to calculate
and the constraint is linear.

	Moreover, as the applications for which our methodologies are devised
require the score calculation to be done once for all or at low
frequency so using a time consuming resolution method is not an issue.

Finally, to calculate a score, it is possible to restrict the relevant
portfolios to involve a low number of risk factors (e.g.~spreads,
involving each only 2 risk factors). The effective dimension of the
exact resolution can then be lowered and the optimization made easier.

\section{Numerical experiments}\label{numerical-experiments}

	We thus performed our experiments on the synthetic Yields curves
provided by the European Central Bank and downloadable at
http://sdw.ecb.europa.eu/:

\begin{itemize}
\itemsep1pt\parskip0pt\parsep0pt
\item
  \textbf{AAA}: synthetic curves aggregated from the $AAA$ issuers of
  the EURO zone (dynamic basket).
\item
  \textbf{ALL}: synthetic curves aggregated from all the issuers of the
  EURO zone (dynamic basket).
\end{itemize}

We used the pillars $ 6M, 1Y, 2Y, 3Y, 4Y, 5Y$ of those yield curves.

	We assume the following setting:

\begin{itemize}
\itemsep1pt\parskip0pt\parsep0pt
\item
  \textbf{Reference set of portfolios}: we consider spread portfolios of
  the form $(B_i, -B_j)$ where:

  \begin{itemize}
  \itemsep1pt\parskip0pt\parsep0pt
  \item
    $B_i$, $B_j$ are some Bonds with semi-annual coupons, with a
    time-to-maturity equal to one of the pillars' maturity
  \item
    $\beta=-1$ and $\beta=-D_i/D_j$.
  \item
    $D_i$, $D_j$ are the durations of the bonds $B_i$ and $B_j$.
  \item
    we obtain $2\times 2 \times\left(\frac{12\times13}{2}\right) = 264$
    portfolios.
  \end{itemize}
\item
  \textbf{Distribution assumption}: a meta-t distribution on the Yield
  rate returns.
\item
  \textbf{Bond pricing}: we approximate the P\&L for a bond to be
  $(\Delta Y)D$ where $\Delta Y$ is the Yield rate move and $D$ the base
  bond duration.
\end{itemize}

	We will obtain the optimal scenarios via numerical optimization.

	\subsection{Low Dimensional or Full Risk
setting}\label{low-dimensional-or-full-risk-setting}

	Should we fit a single distribution on all the risk factor at the same
time, or one on each single test portfolio? While for Gaussian
distribution this does not have an impact, in the case of
meta-elliptical distributions that we are considering, the situation is
a little bit more complicated. This is because, while the marginal
distributions are fixed, the copula can vary.

	From a stability point of view, our decision makes the scores dependent
on the number of risk factors chosen, as the copula is fitted on each
group separately. However, we believe that these additional degrees of
freedom allow to better measure the risk and give a better understanding
of the differences between sets of stress scenarios.

	\subsection{The Stress Scenarios}\label{the-stress-scenarios}

	We consider 2 sets of stress scenarios: a base and an enriched one. As
the scenario generation methodology is not the focus of this work we
decided to use over-simplified and stylized sets. Moreover, this choice
allows us to better highlight the contribution of our scores, as the
difference between the sets havealso a clear interpretation.

	Both the base and the enriched set are obtained starting from the first
three components of a Principal Component Analysis performed separately
on the returns of the \textbf{AAA} and \textbf{ALL} curves. The vectors
are rescaled by a factor $3\times \sigma_i$ where $\sigma_i$ is the
explained standard deviation associated and combined as follows:

\begin{itemize}
\itemsep1pt\parskip0pt\parsep0pt
\item
  The \textbf{base set} $\mathcal{S}$ considers only combination of the
  same level, and with the same sign: $(\pm n^{th}$ component
  \textbf{AAA}, $\pm n^{th}$ component \textbf{ALL}), $n = 1,2,3$, for a
  total 6 possible stressed scenarios;
\item
  The \textbf{enriched set} $\mathcal{S}'$ considers the scenarios in
  the base set, plus the combination given by $(\pm n^{th}$ component
  \textbf{AAA}, $\mp n^{th}$ component \textbf{ALL}), $n = 2,3$, for a
  total 10 possible stressed scenarios.
\end{itemize}

	We plot here the three drivers of the risk scenarios for the two curves.
Notice that the PCA analysis provides three main directions that indeed
qualitatively correspond to the shift, slope change and curvature change
(displaying respectively the sign patterns +, −/+ and +/−/+) often used
in devising hypothetical scenarios for rate curves.

    \begin{center}
    \adjustimage{max size={0.9\linewidth}{0.9\paperheight}}{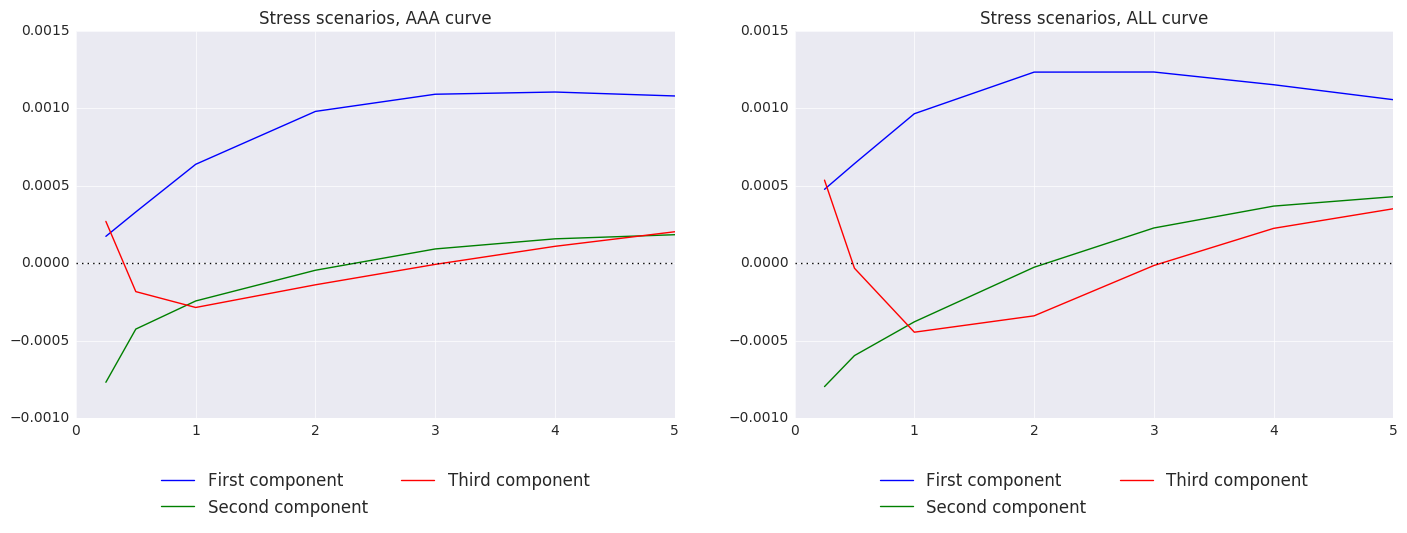}
    \end{center}
    
	\subsection{Result analysis}\label{result-analysis}

	In the table below we compare the average and standard deviation for the
loss to plausibility $\phi_{\cdot}(P)$ and the geometrical score
$\psi_{\cdot}(P)$ aggregated at the driving scenario level, and
considered as a whole (last line). The columns \emph{Quantity} indicates
for how many portfolios the selected scenario produces the largest
losses.


    \begin{center} 
 \begin{tabular}{lrrrrrrrrrr}
\toprule
{} & \multicolumn{5}{l}{Base Scenarios} & \multicolumn{5}{l}{Enriched Scenarios} \\
{Scenario} &       Quantity & $\phi$ mean & $\phi$ std & $\psi$ mean & $\psi$ std &           Quantity & $\phi$ mean & $\phi$ std & $\psi$ mean & $\psi$ std \\
\midrule
(+1, +1) &           80 &    0.334 &   0.258 &    0.858 &   0.193 &               64 &    0.929 &   0.085 &    0.397 &   0.248 \\
(-1, -1) &           80 &    0.348 &   0.263 &    0.872 &   0.171 &               64 &    0.934 &   0.079 &    0.414 &   0.250 \\
(+2, +2) &           28 &    0.442 &   0.244 &    0.512 &   0.331 &               17 &    0.537 &   0.324 &    0.381 &   0.256 \\
(-2, -2) &           28 &    0.437 &   0.246 &    0.497 &   0.343 &               17 &    0.522 &   0.330 &    0.374 &   0.254 \\
(+3, +3) &           24 &    0.666 &   0.259 &    0.746 &   0.274 &               21 &    0.733 &   0.285 &    0.640 &   0.261 \\
(-3, -3) &           24 &    0.658 &   0.262 &    0.736 &   0.283 &               21 &    0.723 &   0.293 &    0.632 &   0.263 \\
(+2, -2) &            0 &    0 &   0 &    0 &   0 &               23 &    0.868 &   0.218 &    0.797 &   0.236 \\
(-2, +2) &            0 &    0 &   0 &    0 &   0 &               23 &    0.869 &   0.225 &    0.802 &   0.237 \\
(+3, -3) &            0 &    0 &   0 &    0 &   0 &                7 &    0.879 &   0.131 &    0.839 &   0.142 \\
(-3, +3) &            0 &    0 &   0 &    0 &   0 &                7 &    0.867 &   0.148 &    0.834 &   0.143 \\
Total    &          264 &    0.420 &   0.285 &    0.766 &   0.282 &              264 &    0.530 &   0.301 &    0.833 &   0.243 \\
\bottomrule
\end{tabular}
 \captionof{table}{Comparison of the scores on the different scores obtained with the different stress scenarios sets.} 
 \end{center}

	We can see that the introduction of the new scenarios does not
deteriorate significantly the score of the existing scenarios (it
actually improves it in some cases), and that the new scenarios have
average scores quite elevated.

It is up to the risk manager to decide if, according to his/her
expertise, the new scenarios are acceptable or not. Particular attention
should be paid in case the scores of the new scenarios are high, but the
average score for some of the old scenarios has been lowered. Again, we
would like to highlight that our scores are not to be intended as an
acceptance tool, but more as a non rejection one.

	We have compared above the score functions $\psi$ and $\phi$ at a
scenario level. In the figure below we compare the two scores at a
portfolio level.

\begin{itemize}
\itemsep1pt\parskip0pt\parsep0pt
\item
  \textbf{Left graph}: the functions $P\rightarrow\phi_{\mathcal{S}}(P)$
  (blue), and $P\rightarrow\phi_{\mathcal{S}'}(P)$ (green). The two
  functions are plotted with the test portfolios in the ascending order
  for $P\rightarrow\phi_{\mathcal{S}}(P)$.
\item
  \textbf{Right graph}: the functions
  $P\rightarrow\psi_{\mathcal{S}}(P)$ (blue), and
  $P\rightarrow\psi_{\mathcal{S}'}(P)$ (green). The two functions are
  plotted with the test portfolios in the ascending order for
  $P\rightarrow\psi_{\mathcal{S}}(P)$.
\end{itemize}

    \begin{center}
    \adjustimage{max size={0.9\linewidth}{0.9\paperheight}}{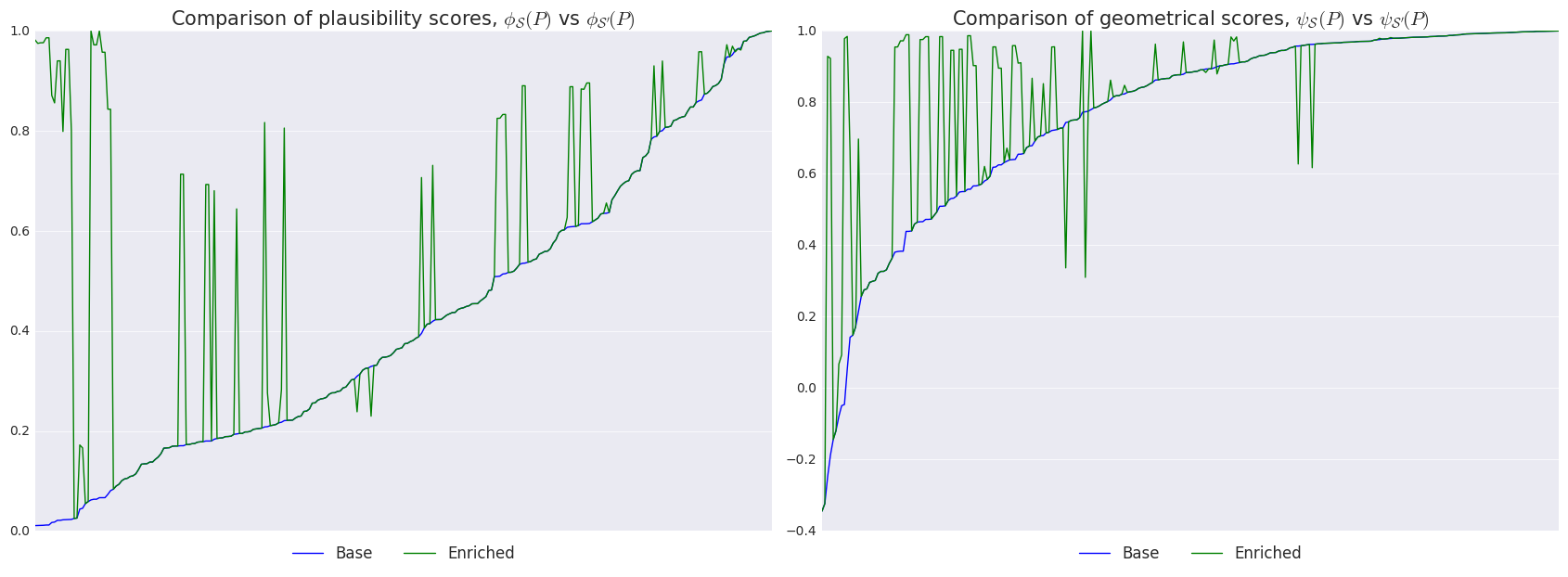}
    \end{center}
    
	We can see that, with the exception of very few portfolios, the scores
obtained by the enriched scenarios are higher than the one obtained by
the base set (when the score is the same the scenario driving the
stressed value is the same in the two sets).

	\subsection{Comparison of probability}\label{comparison-of-probability}

	One natural question is whether or not the new scenarios, even if
obtaining better scores for the analyzed portfolios, are plausible
enough.

A priori some of the new scenarios may be very close or even coincide
with their corresponding most plausible scenarios, however, when
compared with the scenarios previously available they may be way less
plausible. A typical example could be a rescaling of one of the existing
scenarios by a factor $>1$. This could induce (for some test portfolios)
bigger losses, but at the same time the stress scenarios would be less
plausible.

A trade-off between scenario plausibility and losses may happen and it
is up to the risk manager to analyze it and decide if it is acceptable
or not, but it can also happen that the new drivers not only generate
bigger losses but are also more probable, simply because they explore
new direction with respect to the old ones and result in a more
significant position with respect to the reference portfolios.

In the figure below we present the two cases:

\begin{itemize}
\itemsep1pt\parskip0pt\parsep0pt
\item
  \textbf{Left Graph}: Long one bond \textbf{AAA} with maturity $3Y$,
  short one \textbf{ALL} with maturity $3Y$. In this case the new driver
  not only provides a higher loss but also has a higher probability.
\item
  \textbf{Right Graph}: Long one bond \textbf{ALL} with maturity $6M$,
  short one \textbf{AAA} with maturity $6M$. In this case the new driver
  there is a trade-off higher losses lower plausibility.
\end{itemize}

The dashed lines represent the level line of the distribution on the
scenarios (either driver or optimal).

    \begin{center}
    \adjustimage{max size={0.9\linewidth}{0.9\paperheight}}{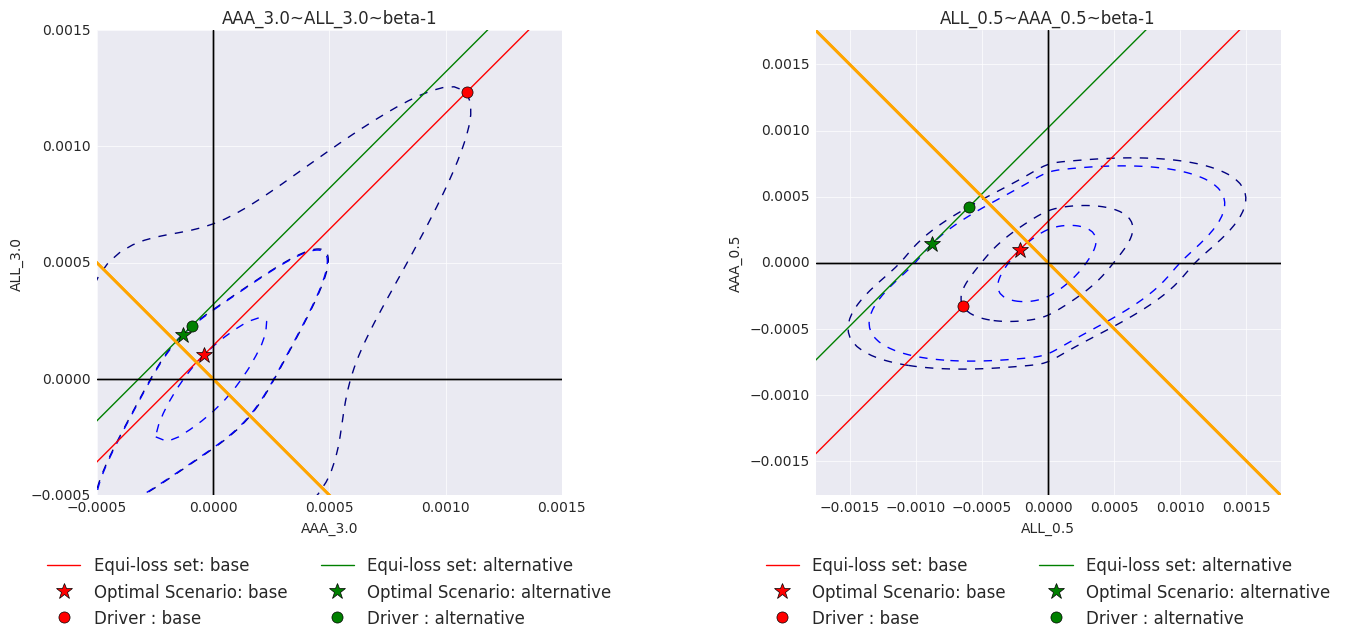}
    \end{center}
    
	The graph on the left shows the most interesting case from a risk
manager point of view. The original set of scenario did not tackle in a
good way the risk associated with the specific test portfolio, as the
driver is orientated almost perpedicularly with respect to the optimal
scenario. The enlarged set introduces a scenario which is in a ``good
direction'' respect to this specific test portfolio, providing a larger
loss ($\approx 2.2$ times the original one) while being simultaneously
many times more plausible (the density ratio between the two driving
scenarios is around $19$).

The graph on the right shows a less appealing case: the enlarged set
introduces a scenario which generates bigger losses but which is, at the
same time, more unlikely (even if both scores are higher in this case).
The risk manager will thus have to decide, based on his expertise, if
the trade off plausibility vs.~risk is acceptable or if the new scenario
(or set of scenarios) should not be taken in consideration.

\section{Conclusions}\label{conclusions}

	In this work we have presented two methodologies which can help risk
managers to compare sets of stress scenarios and in particular to assess
the benefits of the introduction of new scenarios the existing ones. The
two methodologies allow the risk manager to analyze different aspects of
the stress scenarios, notably their position and relevance for the
reference sets of portfolios.

The proposed methodologies have a clear and natural meaning which allows
to better understand the benefit of one set of scenarios with respect to
the other.


    \end{document}